\documentclass[aps,preprint]{revtex4}%
\usepackage{amsfonts}
\usepackage{amsmath}
\usepackage{amssymb}
\usepackage{graphicx}%
\providecommand{\U}[1]{\protect\rule{.1in}{.1in}}
\begin{document}
\title[ ]{A Charged Particle Must be Treated Relativistically in Classical Theory}
\author{Timothy H. Boyer}
\affiliation{Department of Physics, City College of the City University of New York, New
York, New York 10031, USA}
\affiliation{E-mail:  tboyer@ccny.cuny.edu}

\begin{abstract}
A charged particle which is allowed to accelerate must have relativistic
behavior because it is coupled to electromagnetic radiation which propagates
at the speed of light. \ We treat the simple steady-state situation of a
charged particle moving in a circular orbit with counter-propagating plane
waves providing the power which balances the energy radiated away by the
accelerating charge. \ It is emphasized that only an electromagnetic
arrangement, such as a Coulomb potential or a constant magnetic field, can
provide the relativistic central force for the particle motion. \ 

\end{abstract}
\maketitle

\section{Introduction}

\subsection{An Accelerating Charged Particle Must Be Treated Relativistically}

A classical charged particle which is allowed to accelerate must have
relativistic behavior because it is coupled to electromagnetic radiation which
propagates at the speed of light. \ If the charged particle does not have
relativistic behavior, then it cannot be described in a manner which is
independent of the choice of inertial frame. \ Here we consider an example of
this connection between charged particle motion and electromagnetic radiation.
\ We treat the simple steady-state situation of a charged particle moving in a
circular orbit with counter propagating plane waves providing the power which
balances the energy radiated away by the accelerating charge. \ We emphasize
that only an electromagnetic arrangement such as a Coulomb potential or a
constant magnetic field, can provide the relativistic central force required
for the particle motion. \ The textbook examples which use relativistic
mechanical energy for the charged particle while allowing general
nonrelativistic potentials for the particle motion give a false impression to
students. \ 

\subsection{When is a System Relativistic?}

Although it is widely agreed among physicists that Nature is locally
relativistic to a good approximation, relativistic physics is commonly
regarded as needed only for particles of high speed, where one introduces the
relativistic expressions for particle momentum and energy. \ This
high-velocity situation is treated for point-particle collisions in our
elementary classes on relativity and modern physics.\cite{coll} \ However, if
classical particles carry electric charge, then they are immediately connected
to classical electromagnetic radiation which propagates at speed $c$ in vacuum
and hence \textit{must} be treated relativistically. \ Indeed, if any
spatially-extended potential at all exists between particles (charged or
uncharged) in a relativistic theory, then one must turn to a field theory
according to the no-interaction theorem of Currie, Jordan, and
Sudarshan.\cite{noint} \ However, excellent textbooks present particles with
relativistic mechanical momentum in arbitrary nonrelativistic classical
potentials and allow the false impression that these are relativistic systems
when they are not.\cite{mech}

Relativistic systems deal with point particles. \ However, for simple systems,
the relativistic character may not be indicated until a transformation is
made. \ Thus, a free particle moves with constant velocity in \textit{both}
relativistic and nonrelativistic physics. \ Similarly a circular orbit does
not indicate whether or not the centripetal force satisfies relativistic
transformations. \ Only when we carry out a Lorentz transformation to a new
inertial frame, does the relativistic or nonrelativistic character of the
mechanical system become evident. \ \ 

\subsection{Independence of Physical Behavior from the Choice of Inertial
Frame}

One of the foundations of both nonrelativistic and relativistic physics is the
assumption that an observer in any inertial frame can give a description of
the physics which is equivalent to that provided by another observer in a
different inertial frame. \ If the system allows a \textit{relativistic}
description, then the analysis must be consistent in any two inertial frames
connected by Lorentz transformation. In this article, our simple example shows
that the centripetal force on a particle must have \textit{relativistic}
transformation properties to fit consistently with \textit{relativistic}
electromagnetic radiation. \ The use of nonrelativistic physics for the
orbital motion is inconsistent with coupling to relativistic electromagnetic
radiation. \ 

It was pointed out a few years ago that we could consider a collision of two
uncharged particles with mixed transformation properties, treating one
particle using nonrelativistic physics and the second using relativistic
physics.\cite{B2009} \ For such a mixed system, the first three conservation
laws of energy, momentum, and angular momentum can still hold true in any
chosen inertial frame. \ However, the fourth conservation laws (involving
constant motion of the center of mass or involving constant motion of the
center of energy) hold true in no inertial frame for such a mixture. \ Most
importantly, the outcome of the collision depends explicitly upon the inertial
frame in which the energy and momentum conservation laws are applied. \ Thus a
particle-collision system involving the combination of \ a nonrelativistic
system with a relativistic system is dependent upon the inertial frame in
which it is analyzed and does not give a result holding in every inertial frame.

\section{A Charged Particle in Steady-State Motion in an Inertial Frame}

\subsection{Circular Orbit in a Central Potential}

We consider a charged particle of mass $m_{0}$, charge $e$, in a circular
orbit of radius $R$ centered on the origin in the $xy$-plane of inertial frame
$S$,
\begin{equation}
\mathbf{r}\left(  t\right)  =R\left(  \widehat{x}\cos\omega t+\widehat{y}%
\sin\omega t\right)
\end{equation}
due to a central potential
\begin{equation}
\mathcal{V}\left(  r\right)  =\frac{\kappa r^{n}}{n}%
\end{equation}
where $\kappa$ is a positive constant and $n\neq0$. \ Also, there is a uniform
magnetic field $\mathbf{B}=\widehat{z}B_{0}$ in the $z$-direction
perpendicular to the plane of the orbit. \ The radial equation of motion for a
relativistic charged particle corresponding to centripetal acceleration $a$
due to the forces of both the central potential and the magnetic field is
given by%
\begin{equation}
m_{0}\gamma a=m_{0}\gamma\left(  \omega^{2}R\right)  =\kappa r^{n-1}%
-\frac{e\omega R}{c}B_{0}, \label{Newton}%
\end{equation}
where the ratio of speed $\omega R=v$ to the speed of light $c$ is $\omega
R/c=v/c=\beta$ and $\gamma=\left(  1-\beta^{2}\right)  ^{-1/2}$. \ 

\subsection{Energy-Balancing Radiation}

If the particle is charged, it is immediately connected thorough its charge to
electromagnetic fields. \ Because the relativistic charged particle is
accelerating, it is emitting radiation energy at a rate\cite{Jackson2}%

\begin{equation}
P_{emitted}=\frac{2e^{2}}{3c^{3}}\gamma^{6}c^{2}\left[  \dot{\beta}%
^{2}-\left(  \mathbf{\beta\times\dot{\beta}}\right)  ^{2}\right]
=\frac{2e^{2}}{3c^{3}}\gamma^{4}\omega^{4}R^{2}. \label{rad}%
\end{equation}
The relativistic power lost involves a factor of $\gamma^{4}$ multiplying the
familiar nonrelativistic Larmor formula for a circular orbit. \ 

The particle can remain in circular motion in steady state despite its
emission of radiation if it is provided energy by electromagnetic radiation
fields. \ Here we assume the simple situation of two counter-propagating
circularly-polarized plane waves. \ The wave traveling in the plus-$z$
direction is given by%

\begin{equation}
\mathbf{E}_{+}(z,t)=\frac{E_{0}}{2}\left[  -\widehat{x}\sin\left(  -Kz+\Omega
t\right)  +\widehat{y}\cos\left(  -Kz+\Omega t\right)  \right]  ,
\end{equation}%
\begin{equation}
\mathbf{B}_{+}(z,t)=\frac{E_{0}}{2}\left[  -\widehat{y}\sin\left(  -Kz+\Omega
t\right)  -\widehat{x}\cos\left(  -Kz+\Omega t\right)  \right]  ,
\end{equation}
and that in the minus-$z$ direction is \qquad%
\begin{equation}
\mathbf{E}_{-}(z,t)=\frac{E_{0}}{2}\left[  -\widehat{x}\sin\left(  Kz+\Omega
t\right)  +\widehat{y}\cos\left(  Kz+\Omega t\right)  \right]  ,
\end{equation}%
\begin{equation}
\mathbf{B}_{-}(z,t)=\frac{E_{0}}{2}\left[  \widehat{y}\sin\left(  Kz+\Omega
t\right)  +\widehat{x}\cos\left(  Kz+\Omega t\right)  \right]  ,
\end{equation}
where $\Omega=cK$. \ Then in the $xy$-plane where $z=0$, the electric
radiation field is
\begin{equation}
\mathbf{E}\left(  0,t\right)  =\mathbf{E}_{+}(0,t)+\mathbf{E}_{-}%
(0,t)=E_{0}\left[  -\widehat{x}\sin\left(  \Omega t\right)  +\widehat{y}%
\cos\left(  \Omega t\right)  \right]  , \label{Esum}%
\end{equation}
while the magnetic field due to the radiation vanishes in the plane
\begin{equation}
\mathbf{B}\left(  0,t\right)  =0. \label{Bsum}%
\end{equation}

In order to provide a steady-state situation between the particle motion and
the radiation, we require that the frequencies agree, $\omega=\Omega.$
\ Furthermore, we require the amplitude $E_{0}$ of the electric radiation
fields is such as to balance the energy radiated away by the accelerating
charged particle in its circular orbit. \ Thus from Eq. (\ref{rad}), we
require the power balance%
\begin{equation}
P_{absorbed}=e\omega RE_{0}=\frac{2e^{2}}{3c^{3}}\gamma^{4}\omega^{4}%
R^{2}=P_{emitted}. \label{Ubal}%
\end{equation}

\section{Lorentz Transformation to a New Inertial Frame $S^{\prime}$}

\subsection{The $S^{\prime}$ Inertial Frame}

Now we consider this same physical system as seen in a new inertial frame
$S^{\prime}$ obtained by a Lorentz transformation with relative velocity
$\mathbf{V=}\widehat{z}V$ relative to the initial inertial frame $S$. \ Thus
$S$ labels the original inertial frame where the particle orbit is in the
$xy$-plane, and $S^{\prime}$ labels the new inertial frame in which the
$z$-component of particle velocity is $v_{z}=-V$. \ The Lorentz
transformations are exactly those of junior-level electromagnetism
texts.\cite{Griffiths}

In the $S^{\prime}$ inertial frame, the radius of the orbit is still $R$, but
the orbit is now given by
\begin{equation}
\mathbf{r}^{\prime}\left(  t^{\prime}\right)  =R\left(  \widehat{x}\cos
\omega^{\prime}t^{\prime}+\widehat{y}\sin\omega^{\prime}t^{\prime}\right)
-\widehat{z}Vt^{\prime} \label{rprime}%
\end{equation}
with velocity%
\begin{equation}
\mathbf{v}^{\prime}\left(  t^{\prime}\right)  =\mathbf{\dot{r}}^{\prime
}\left(  t^{\prime}\right)  =\omega^{\prime}R\left(  -\widehat{x}\sin
\omega^{\prime}t^{\prime}+\widehat{y}\cos\omega^{\prime}t^{\prime}\right)
-\widehat{z}V \label{vprime}%
\end{equation}
and acceleration toward the $z$-axis
\begin{equation}
\mathbf{a}^{\prime}\left(  t^{\prime}\right)  =\mathbf{\ddot{r}}^{\prime
}\left(  t^{\prime}\right)  =-\left(  \omega^{\prime}\right)  ^{2}R\left(
\widehat{x}\cos\omega^{\prime}t^{\prime}+\widehat{y}\sin\omega^{\prime
}t^{\prime}\right)  , \label{aprime}%
\end{equation}
where $t^{\prime}=\Gamma\left[  t-\left(  V/c\right)  0\right]  =\Gamma t$,
and the orbital frequency in $S^{\prime}$ is $\omega^{\prime}=\omega/\Gamma$
with $\Gamma=\left(  1-V^{2}/c^{2}\right)  ^{-1/2}$. \ We notice that in the
$S^{\prime}$ frame, the frequency $\omega^{\prime}=\omega/\Gamma$ is smaller
than the frequency $\omega$ in $S$, consistent with the slowing down of moving
clocks. \ 

\subsection{Particle Motion in $S^{\prime}$}

Now we want to check Newton's second law for the particle motion in the
$S^{\prime}$ frame. \ In the $S^{\prime}$ inertial frame, the particle speed
has both a radial component and a $z$-component so that its speed is
$v^{\prime}=\sqrt{\left(  \omega^{\prime}R\right)  ^{2}+V^{2}}$, and its gamma
factor is $\gamma^{\prime}=\left\{  1-\left[  \left(  \omega^{\prime}R\right)
^{2}+V^{2}\right]  /c^{2}\right\}  ^{-1/2}$. \ In the centripetal direction,
we require%
\begin{align}
m_{0}\gamma^{\prime}a^{\prime}  &  =m_{0}\gamma^{\prime}\left[  \left(
\omega^{\prime}\right)  ^{2}R\right]  =\left(  \kappa r^{n-1}\right)
^{\prime}-\frac{e}{c}v^{\prime}B_{0}\nonumber\\
&  =\left(  \kappa r^{n-1}\right)  ^{\prime}-\frac{e}{c}\omega^{\prime}RB_{0},
\end{align}
where $\left(  \kappa r^{n-1}\right)  ^{\prime}$ is the centripetal force in
the $S^{\prime}$ inertial frame. \ 

\subsubsection{Case of Cyclotron Motion}

First let us consider the case of centripetal motion involving only a magnetic
field in the $z$-direction. \ Then in the $S$ frame, we have $\gamma=\left(
1-\omega^{2}R^{2}/c^{2}\right)  ^{-1/2}$ and the equation of motion
(\ref{Newton}) becomes
\begin{equation}
m_{0}\gamma\left(  \omega^{2}R\right)  =\frac{-e}{c}\omega RB_{0}.
\label{cycl}%
\end{equation}
In the $S^{\prime}$ frame, we have the radial equation as%
\begin{equation}
m_{0}\gamma^{\prime}\left[  \left(  \omega^{\prime}\right)  ^{2}R\right]
=\frac{-e}{c}\omega^{\prime}RB_{0}. \label{cyclp}%
\end{equation}
The gamma factor $\gamma^{\prime}$ can be rewritten as
\begin{align}
\gamma^{\prime}  &  =\left\{  1-\left[  \left(  \omega^{\prime}R\right)
^{2}+V^{2}\right]  /c^{2}\right\}  ^{-1/2}\nonumber\\
&  =\left\{  1-\frac{\left(  \omega R\right)  ^{2}}{c^{2}}\left(
1-\frac{V^{2}}{c^{2}}\right)  -\frac{V^{2}}{c^{2}}\right\}  ^{-1/2}\nonumber\\
&  =\left(  1-\frac{V^{2}}{c^{2}}\right)  ^{-1/2}\left(  1-\frac{\left(
\omega R\right)  ^{2}}{c^{2}}\right)  ^{-1/2}=\Gamma\gamma. \label{gammap}%
\end{align}
Also, we have $\omega^{\prime}=\omega/\Gamma$. \ But then the radial equation
of motion (\ref{cyclp}) in $S^{\prime}$ becomes
\begin{equation}
m_{0}\gamma^{\prime}\left[  \left(  \omega^{\prime}\right)  ^{2}R\right]
=m_{0}\Gamma\gamma\left[  \left(  \frac{\omega}{\Gamma}\right)  ^{2}R\right]
=\frac{m_{0}\gamma\left[  \omega^{2}R\right]  }{\Gamma}=\frac{-e\omega RB_{0}%
}{c\Gamma}. \label{gammapp}%
\end{equation}
Canceling the common factor of $1/\Gamma$ between the last two terms in Eq.
(\ref{gammapp}), we find the same condition as appears in Eq. (\ref{cycl}) in
the $S$ inertial frame. \ Thus, Newton's second law is treated consistently in
both inertial frames. \ 

\subsubsection{Coulomb Central Potential}

Next consider a purely Coulomb potential $\mathcal{V}(r)=-Ze^{2}/r$ providing
the central force. \ In the $S$ frame, there is only an electrostatic Coulomb
field due to the nucleus and no magnetic field. \ Thus in the $S$ frame, we
have $\gamma=\left[  1-\omega^{2}R^{2}/c^{2}\right]  ^{-1/2}$ and
\begin{equation}
m_{0}\gamma\omega^{2}R=\frac{Ze^{2}}{R^{2}}. \label{CoulS}%
\end{equation}

In the $S^{\prime}$ inertial frame, we find both electric and magnetic fields
arising from the Coulomb potential in $S$. \ Thus we can think of the Coulomb
potential as arising from an infinitely massive (nuclear) point charge $-Ze$
at the coordinate origin in $S$. \ In the $S^{\prime}~$frame, the radial
equation becomes
\begin{equation}
m_{0}\gamma^{\prime}\left[  \left(  \omega^{\prime}\right)  ^{2}R\right]
=-eE_{r}^{\left(  N\right)  \prime}-\frac{e}{c}(-V)B_{\phi}^{\left(  N\right)
\prime} \label{CoulSp}%
\end{equation}
where $E_{r}^{\left(  N\right)  \prime}$ and $B_{\phi}^{\left(  N\right)
\prime}$ are the components of the electromagnetic fields in $S^{\prime}$ due
to the (nuclear) charge $-Ze$ evaluated at the position of the charge $e$.
\ As seen in the $S^{\prime}$ inertial frame, the Coulomb center (the nucleus
with charge $-Ze$) is moving with constant velocity $-V$ in the $z$-direction.
\ The force on the charge $e$ can be regarded as arising from retarded fields
due to the charge $-Ze,$ or can be obtained from Lorentz transformation of the
fields of a charge moving with constant velocity. \ The fields of the Coulomb
center $-Ze$ moving with constant velocity in the $S^{\prime}$ inertial frame
are\cite{Griffiths}
\begin{equation}
\mathbf{E}^{\prime}(\mathbf{r}^{\prime},t^{\prime})=-Ze\Gamma\frac{\left[
\widehat{x}x^{\prime}+\widehat{y}y^{\prime}+\widehat{z}\left(  z^{\prime
}+Vt^{\prime}\right)  \right]  }{\left[  x^{^{\prime}2}+y^{\prime2}+\Gamma
^{2}\left(  z^{\prime}+Vt^{\prime}\right)  ^{2}\right]  ^{3/2}}%
\end{equation}
and%
\begin{equation}
\mathbf{B}^{\prime}(\mathbf{r}^{\prime},t^{\prime})=-Ze\frac{V}{c}\Gamma
\frac{\left[  \widehat{y}x^{\prime}-\widehat{x}y^{\prime}\right]  }{\left[
x^{^{\prime}2}+y^{\prime2}+\Gamma^{2}\left(  z^{\prime}+Vt^{\prime}\right)
^{2}\right]  ^{3/2}}.
\end{equation}
Substituting from Eq. (\ref{rprime}) and noting $z^{\prime}=-Vt^{\prime},$ we
obtain the electromagnetic fields due to the (nucleus) $-Ze$ at the position
of the charge $e$ in its circular orbit of radius $R$ as
\[
E_{r}^{\left(  N\right)  \prime}=-Ze\Gamma\frac{1}{R^{2}}%
\]
and%
\[
B_{\phi}^{\left(  N\right)  \prime}=-Ze\frac{\left(  -V\right)  }{c}%
\Gamma\frac{1}{R^{2}}.
\]
But then the radial equation of motion for the orbit in the $S^{\prime}$
inertial frame involves an electric force and also a magnetic force arising
from the $z^{\prime}$-component of the particle's velocity,%
\begin{align}
-eE_{r}^{\left(  N\right)  \prime}-\frac{e}{c}\left(  -V\right)  B_{\phi
}^{\left(  N\right)  \prime}  &  =e\left(  Ze\Gamma\frac{1}{R^{2}}\right)
+\frac{e}{c}\left(  -V\right)  \left(  Ze\frac{\left(  V\right)  }{c}%
\Gamma\frac{1}{R^{2}}\right) \nonumber\\
&  =\Gamma\left(  Ze^{2}\frac{1}{R^{2}}\right)  \left(  1-\frac{V^{2}}{c^{2}%
}\right)  =\frac{1}{\Gamma}\left(  Ze^{2}\frac{1}{R^{2}}\right)  ,
\label{FCoul}%
\end{align}
since $\Gamma=\left(  1-V^{2}/c^{2}\right)  ^{-1/2}$. \ Thus, from equations
(\ref{CoulSp}) and (\ref{FCoul}), we have%
\begin{equation}
m_{0}\gamma^{\prime}\left[  \left(  \omega^{\prime}\right)  ^{2}R\right]
=\frac{m_{0}\gamma\left[  \omega^{2}R\right]  }{\Gamma}=\frac{1}{\Gamma}%
\frac{Ze^{2}}{R^{2}}. \label{CoulSpp}%
\end{equation}
Again, we can cancel the common factor of $1/\Gamma$ in the last two terms in
Eq. (\ref{CoulSpp}) and so obtain Eq. (\ref{CoulS}) which held in the $S$
inertial frame. \ Again, there is consistency between the $S$ and $S^{\prime}$
inertial frames. \ 

\subsection{Radiation Energy Balance in $S^{\prime}$}

Although we have now shown that the Lorentz transformation from the $S$ to the
$S^{\prime}$ frame gives consistent radial equations of motion for both
cyclotron motion and motion in a Coulomb potential, we must now show
consistency for the energy balance associated with radiation emission and absorption.

\subsubsection{Radiation Emission}

In the $S^{\prime}$ inertial frame, the power emitted by the orbiting charge
is given by\cite{Jackson2} $P^{\prime}=\left[  2e^{2}/\left(  3c\right)
\right]  \gamma^{^{\prime}6}\left[  \left(  \dot{\beta}^{\prime}\right)
^{2}-\left(  \mathbf{\beta}^{\prime}\mathbf{\times\dot{\beta}}^{\prime
}\right)  ^{2}\right]  $. \ From equations (\ref{vprime}) and (\ref{aprime}),
we will need
\begin{equation}
\dot{\beta}^{\prime}=a^{\prime}/c=\left(  \omega^{\prime}\right)
^{2}R/c=\frac{\omega^{2}R}{c}\left(  1-\frac{V^{2}}{c^{2}}\right)
\end{equation}
and
\begin{align}
\mathbf{\beta}^{\prime}\mathbf{\times\dot{\beta}}^{\prime}  &  =\frac{1}%
{c^{2}}\mathbf{v}^{\prime}\times\mathbf{a}^{\prime}\nonumber\\
&  =\frac{1}{c^{2}}\left[  \omega^{\prime}R\left(  -\widehat{x}\sin
\omega^{\prime}t^{\prime}+\widehat{y}\cos\omega^{\prime}t^{\prime}\right)
-\widehat{z}V\right]  \times\left[  -\left(  \omega^{\prime}\right)
^{2}R\left(  \widehat{x}\cos\omega^{\prime}t^{\prime}+\widehat{y}\sin
\omega^{\prime}t^{\prime}\right)  \right] \nonumber\\
&  =\frac{-\left(  \omega^{\prime}\right)  ^{2}R}{c^{2}}\left[  \widehat{x}%
\left(  V\sin\omega^{\prime}t^{\prime}\right)  +\widehat{y}\left(
-V\cos\omega^{\prime}t^{\prime}\right)  -\widehat{z}\omega^{\prime}R\right]  ,
\end{align}
giving
\begin{align}
&  \left(  \dot{\beta}^{\prime}\right)  ^{2}-\left(  \mathbf{\beta}^{\prime
}\mathbf{\times\dot{\beta}}^{\prime}\right)  ^{2}\nonumber\\
&  =\left[  \frac{\omega^{2}R}{c}\left(  1-\frac{V^{2}}{c^{2}}\right)
\right]  ^{2}-\frac{\left(  \omega\right)  ^{4}R^{2}\left(  1-V^{2}%
/c^{2}\right)  ^{2}}{c^{2}}\left[  \frac{V^{2}}{c^{2}}+\frac{\left(
\omega\right)  ^{2}R^{2}}{c^{2}}\left(  1-V^{2}/c^{2}\right)  \right]
\nonumber\\
&  =\left[  \frac{\omega^{2}R}{c}\left(  1-\frac{V^{2}}{c^{2}}\right)
\right]  ^{2}\left(  1-\frac{V^{2}}{c^{2}}\right)  \left(  1-\frac{\left(
\omega\right)  ^{2}R^{2}}{c^{2}}\right)  =\frac{\omega^{4}R^{2}}{c^{2}}%
\Gamma^{-6}\gamma^{-2}%
\end{align}
Then in the $S^{\prime}$ inertial frame, the power radiated is
\begin{align}
P_{emitted}^{\prime}  &  =\frac{2e^{2}}{3c}\left(  \gamma^{\prime}\right)
^{6}\left[  \left(  \dot{\beta}^{\prime}\right)  ^{2}-\left(  \mathbf{\beta
}^{\prime}\mathbf{\times\dot{\beta}}^{\prime}\right)  ^{2}\right] \nonumber\\
&  =\frac{2e^{2}}{3c}\gamma^{6}\Gamma^{6}\left[  \frac{\omega^{4}R^{2}}{c^{2}%
}\Gamma^{-6}\gamma^{-2}\right] \nonumber\\
&  =\frac{2e^{2}}{3c^{3}}\gamma^{4}\omega^{4}R^{2},
\end{align}
which is exactly the result in Eq. (\ref{rad}) for the power radiated in the
$S$ inertial frame. \ Indeed, the power radiated by an accelerating point
charge is independent of the inertial frame in which it is
evaluated.\cite{Jackson2}

\subsection{Plane-Wave Radiation in the $S^{\prime}$ Inertial Frame}

We still need to confirm that the counter propagating plane waves indeed
provide the correct power to the charged particle in the $S^{\prime}$ inertial
frame. \ The circularly-polarized plane waves remain circularly-polarized
plane waves in the $S^{\prime}$ inertial frame,\cite{Griffiths}
\begin{equation}
\mathbf{E}_{+}^{\prime}(z^{\prime},t^{\prime})=\sqrt{\frac{1-V/c}{1+V/c}}%
\frac{E_{0}}{2}\left[  \widehat{x}\sin\left(  -K_{+}^{\prime}z^{\prime}%
+\Omega_{+}^{\prime}t^{\prime}\right)  +\widehat{y}\cos\left(  -K_{+}^{\prime
}z^{\prime}+\Omega_{+}^{\prime}t^{\prime}\right)  \right]  ,
\end{equation}%
\begin{equation}
\mathbf{B}_{+}^{\prime}(z^{\prime},t^{\prime})=\sqrt{\frac{1-V/c}{1+V/c}}%
\frac{E_{0}}{2}\left[  \widehat{y}\sin\left(  -K_{+}^{\prime}z^{\prime}%
+\Omega_{+}^{\prime}t^{\prime}\right)  -\widehat{x}\cos\left(  -K_{+}^{\prime
}z^{\prime}+\Omega_{+}^{\prime}t^{\prime}\right)  \right]  ,
\end{equation}
where
\begin{equation}
cK_{+}^{\prime}=\Omega_{+}^{\prime}=\omega\sqrt{\frac{1-V/c}{1+V/c}},
\end{equation}
and \qquad%
\begin{equation}
\mathbf{E}_{-}^{\prime}(z^{\prime},t^{\prime})=\sqrt{\frac{1+V/c}{1-V/c}}%
\frac{E_{0}}{2}\left[  \widehat{x}\sin\left(  K_{-}^{\prime}z^{\prime}%
+\Omega_{-}^{\prime}t^{\prime}\right)  +\widehat{y}\cos\left(  K_{-}^{\prime
}z^{\prime}+\Omega_{-}^{\prime}t^{\prime}\right)  \right]  ,
\end{equation}%
\begin{equation}
\mathbf{B}_{-}^{\prime}(z^{\prime},t^{\prime})=\sqrt{\frac{1+V/c}{1-V/c}}%
\frac{E_{0}}{2}\left[  -\widehat{y}\sin\left(  K_{-}^{\prime}z^{\prime}%
+\Omega_{-}^{\prime}t^{\prime}\right)  +\widehat{x}\cos\left(  K_{-}^{\prime
}z^{\prime}+\Omega_{-}^{\prime}t^{\prime}\right)  \right]  ,
\end{equation}
where
\begin{equation}
cK_{-}^{\prime}=\Omega_{-}^{\prime}=\omega\sqrt{\frac{1+V/c}{1-V/c}}.
\end{equation}

Then in the particle's orbital plane where $z^{\prime}=-Vt^{\prime}$, we have%
\begin{align}
-K_{+}^{\prime}z^{\prime}+\Omega_{+}^{\prime}t^{\prime}  &  =-\left(
\Omega_{+}^{\prime}/c\right)  \left(  -Vt^{\prime}\right)  +\Omega
_{+}t^{\prime}=\left(  1+V/c\right)  \Omega_{+}t^{\prime}\nonumber\\
&  =\left(  1+V/c\right)  \left(  \omega\sqrt{\frac{1-V/c}{1+V/c}}\right)
t^{\prime}=\omega\sqrt{1-\left(  V/c\right)  ^{2}}t^{\prime}=\frac{\omega
}{\Gamma}t^{\prime}=\omega^{\prime}t^{\prime},
\end{align}
and similarly,
\begin{equation}
K_{-}^{\prime}z^{\prime}+\Omega_{-}^{\prime}t^{\prime}=\left(  1-V/c\right)
\left(  \omega\sqrt{\frac{1+V/c}{1-V/c}}\right)  t^{\prime}=\omega^{\prime
}t^{\prime}.
\end{equation}
Thus the two plane waves are in phase in the orbital plane of the charge with
the sum of the electric fields
\begin{align}
\mathbf{E}^{\prime}\left(  -Vt^{\prime},t^{\prime}\right)   &  =\mathbf{E}%
_{+}^{\prime}(-Vt^{\prime},t^{\prime})+\mathbf{E}_{-}^{\prime}(-Vt^{\prime
},t^{\prime})\nonumber\\
&  =\left(  \sqrt{\frac{1-V/c}{1+V/c}}+\sqrt{\frac{1+V/c}{1-V/c}}\right)
\frac{E_{0}}{2}\left[  -\widehat{x}\sin\left(  \omega^{\prime}t^{\prime
}\right)  +\widehat{y}\cos\left(  \omega^{\prime}t^{\prime}\right)  \right]
\nonumber\\
&  =\frac{1}{\sqrt{1-V^{2}/c^{2}}}E_{0}\left[  -\widehat{x}\sin\left(
\omega^{\prime}t^{\prime}\right)  +\widehat{y}\cos\left(  \omega^{\prime
}t^{\prime}\right)  \right]  , \label{Ep}%
\end{align}
and a sum of the magnetic field contributions,
\begin{align}
\mathbf{B}^{\prime}\left(  -Vt^{\prime},t^{\prime}\right)   &  =\mathbf{B}%
_{+}^{\prime}(-Vt^{\prime},t^{\prime})+\mathbf{B}_{-}^{\prime}(-Vt^{\prime
},t^{\prime})\nonumber\\
&  =\sqrt{\frac{1-V/c}{1+V/c}}\frac{E_{0}}{2}\left[  \widehat{y}\sin\left(
\omega^{\prime}t^{\prime}\right)  -\widehat{x}\cos\left(  \omega^{\prime
}t^{\prime}\right)  \right] \nonumber\\
&  +\sqrt{\frac{1+V/c}{1-V/c}}\frac{E_{0}}{2}\left[  -\widehat{y}\sin\left(
\omega^{\prime}t^{\prime}\right)  +\widehat{x}\cos\left(  \omega^{\prime
}t^{\prime}\right)  \right] \nonumber\\
&  =\frac{-V/c}{\sqrt{1-V^{2}/c^{2}}}E_{0}\left[  \widehat{y}\sin\left(
\omega^{\prime}t^{\prime}\right)  -\widehat{x}\cos\left(  \omega^{\prime
}t^{\prime}\right)  \right]  . \label{Bp}%
\end{align}
Indeed, if we carry out the usual Lorentz transformations for the transverse
electromagnetic fields $\mathbf{E=}\widehat{x}E_{x}+\widehat{y}E_{y}$ and
$\mathbf{B=0}$ in Eqs. (\ref{Esum}) and (\ref{Bsum}), we obtain, using
$\mathbf{E}^{\prime}=\Gamma\mathbf{E}$ and $\mathbf{B}^{\prime}=\left(
\Gamma\mathbf{V}\right)  \times\mathbf{E)}$ for the fields perpendicular to
the relative velocity $\mathbf{V,}$ exactly the results in Eqs. (\ref{Ep}) and
(\ref{Bp}). \ In the direction parallel to the relative velocity between the
frames, the electric and magnetic fields are unchanged so that $E_{z}^{\prime
}=0$ and $B_{z}^{\prime}=B_{z}=B_{0}$. \ 

The power delivered to the orbiting charge by the circularly polarized
radiation fields is
\begin{align}
P_{absorbed}^{\prime}  &  =e\mathbf{E}^{\prime}\left(  \mathbf{r}_{e}^{\prime
},t^{\prime}\right)  \cdot\mathbf{v}_{e}^{\prime}\nonumber\\
&  =e\left\{  \frac{1}{\sqrt{1-V^{2}/c^{2}}}E_{0}\left[  -\widehat{x}%
\sin\left(  \omega^{\prime}t^{\prime}\right)  +\widehat{y}\cos\left(
\omega^{\prime}t^{\prime}\right)  \right]  \right\} \nonumber\\
&  \cdot\left[  \omega^{\prime}R\left(  -\widehat{x}\sin\omega^{\prime
}t^{\prime}+\widehat{y}\cos\omega^{\prime}t^{\prime}\right)  -\widehat{z}%
V\right] \nonumber\\
&  =\left(  e\Gamma E_{0}\right)  \omega^{\prime}R=\left(  e\Gamma
E_{0}\right)  \left(  \omega/\Gamma\right)  R=eE_{0}\omega R,
\end{align}
exactly the same power as absorbed by the charge in the $S$ inertial frame as
given in Eq. (\ref{Ubal}).\ 

\subsubsection{Coulomb Potential and a Magnetic Field $B_{0}$}

There is no problem treating the situation of a circular orbit under the
influence of \textit{both} a Coulomb potential and a magnetic field as in Eq.
(\ref{Newton}). \ The rate of change of relativistic momentum transforms as in
Eq. (\ref{gammapp}) while the forces transform as in Eqs. (\ref{gammapp}) and
(\ref{FCoul}). \ The energy balance remains as in Eq. (\ref{Ubal})

\subsection{General Nonrelativistic Potential}

\subsubsection{\ Force Transformation}

We mentioned earlier that for simple systems, the assumed nonrelativistic or
relativistic character may become apparent only when the systems are
considered in a different inertial frame. \ If we consider a circular orbit
for a charged particle moving in a general central potential in frame $S$, we
have a problem when we view the situation from the $S^{\prime}$ inertial
frame. \ Thus in $S$, we have the (apparently) relativistic centripetal
acceleration equation
\begin{equation}
\widehat{r}\cdot\frac{d\mathbf{p}}{dt}=-m_{0}\gamma\omega^{2}R=-\frac
{\partial\mathcal{V}}{\partial r}.
\end{equation}
The transformation of the rate of change of momentum involves no difficulty
since the relativistic expression has been used. \ Thus, in $S^{\prime}$, we
have just as before%
\begin{equation}
\widehat{r}^{\prime}\cdot\frac{d\mathbf{p}^{\prime}}{dt^{\prime}}=m_{0}%
\gamma^{\prime}\omega^{\prime2}R=m_{0}\Gamma\gamma\left(  \frac{\omega}%
{\Gamma}\right)  ^{2}R=\frac{m_{0}\gamma\omega^{2}R}{\Gamma}.
\end{equation}
However, the Lorentz transformation of a force from a general central
potential $\mathcal{V}\left(  r\right)  $ remains problematic. \ Under a
\textit{nonrelativistic} transformation between inertial frames, a force is
unchanged. \ Indeed, if the central potential takes the nonrelativistic form
$\mathcal{V}\left(  r\right)  =kr^{n}/n$, and we simply transform the
unchanged radial distance nonrelativistically as $r^{\prime}=r=R$, then we
find $\partial\mathcal{V}/\partial r=\partial\mathcal{V}^{\prime}/\partial
r^{\prime}$, which is incorrect since it lacks the factor of $1l\Gamma$ which
appears in the relativistic transformations leading to Eqs. (\ref{gammapp})
and (\ref{CoulSpp}), \ What we require is that the force on the charge
particle in the circular orbit is a Lorentz force and transforms as a Lorentz
force, consistent with the field transformations for the electric and magnetic
fields. \ 

Indeed, only in the case of a central potential corresponding to the Coulomb
potential do we know how to proceed because we may regard the Coulomb
potential as arising from the electromagnetic fields of a massive point charge
$-Ze$, and we know how to transform the electromagnetic fields of a point
charge between inertial frames.

The force on the charged particle producing the circular orbit must transform
as a Lorentz force between inertial frames. \ Our example suggests that the
use of relativistic momentum and energy expressions for a point particle in a
central potential will result in a relativistic situation only if the charged
particle is in a Coulomb potential and/or a magnetic field. \ Only classical
electrodynamics provides the basis for a familiar relativistic classical
theory. \ 

\section{Conclusion}

Classical electrodynamics is a fully relativistic theory. \ The physicists of
the early 20th century assumed incorrectly that relativistic electromagnetic
theory can be combined with arbitrary nonrelativistic mechanics.\cite{bb}
\ Such an assumption is quite clear in Lorentz's \textit{Theory of
Electrons}.\cite{Lorentz} \ Indeed, the erroneous assumption that any
nonrelativistic potential can be combined with relativistic expressions for
mechanical particle energy and momentum to create a fully relativistic system
persists in the research literature.\cite{Blanco} \ Modern textbooks of
mechanics and modern physics, which regard as a relativistic system, particles
with \textit{relativistic} momentum in a general \textit{nonrelativistic}
potential, do not help readers by continuing the same false
impression.\cite{mech}\ \ Despite the facile statement that \textquotedblleft
everyone knows that the only relativistic systems those which have all aspects
fully relativistic,\textquotedblright\ the physics literature suggests that
physicist do not follow this rule. \ Calculations in the literature ignore the
no-interaction theorem and the sort of analysis given in the present
manuscript. \ 

The author has no known conflict of interest.

ChargedParticle2.tex \ \ April 27, 2023
\end{document}